\documentclass[a4,11pt]{article}
\usepackage{authblk}
\usepackage{graphicx}

\usepackage{times}
\usepackage{amsmath}    
\usepackage{amsfonts,bbm,bm}   
\usepackage{amssymb}
\usepackage{graphicx}
\usepackage{overpic}
\usepackage[numbers]{natbib}
\usepackage{subfigure}
\usepackage{longtable}
\usepackage{setspace}
\usepackage{lineno}
\usepackage{color}
\usepackage[utf8]{inputenc}

\usepackage{amsfonts}
\usepackage{amssymb}
\usepackage{multirow}
\usepackage[table,xcdraw]{xcolor}
\usepackage{eurosym}
\usepackage[draft]{changes}
\usepackage{url}
\newcommand{\D}{\mathrm{d}}
\newcommand{\Do}{\mathcal{D}}
\newcommand{\bfs}{{\bf s}}
\newcommand{\bfx}{{\bf x}}
\newcommand{\bfz}{{\bf z}}
\newcommand{\bfr}{{\bf r}}
\newcommand{\rr}{\| \bfr \|}

\newcommand{\st}{\bfs,t}
\newcommand{\stp}{{\bfs',t'}}

\newcommand{\EE}[1]{\mathcal{E}\left[{\, #1} \, \right]}
\newcommand{\bfk}{{\bf k}}
\newcommand{\kk}{\|{\bf k}\|}

\newcommand{\cors}{C_{\rmx}({\bf r},\tau; \bmthe')}
\newcommand{\gam}{\gamma}
\newcommand{\rmx}{{\rm x}}

\newcommand{\E}{\mathrm{e}}
\newcommand{\de}{\delta}
\newcommand{\al}{\alpha}
\newcommand{\be}{\beta}

\newcommand{\Rd}{\mathbb{R}^{d}}
\newcommand{\sign}{\mathrm{sign}}

\newcommand{\Ha}{\mathcal{H}}
\newcommand{\R}{\mathbb{R}}
\newcommand{\Or}{\mathcal{O}}

\newcommand{\noi}{\noindent}
\newcommand{\la}{\lambda}

\newcommand{\beq}{\begin{equation}}
\newcommand{\eeq}{\end{equation}}

\newcommand{\e}{{\eta_1}}
\newcommand{\eo}{{\eta_0}}

\newcommand{\bmthe}{{\bm \theta}}
\providecommand{\keywords}[1]{\textbf{\textit{Keywords:}} #1}
\DeclareMathOperator\erf{erf}
\DeclareMathOperator\erfc{erfc}

\begin{document} 
\title{Space-Time Covariance Functions based on Linear Response Theory and
the Turning Bands Method\thanks{This manuscript is a preprint of an article submitted for publication in Journal of Spatial Statistics, \texttt{http://www.journals.elsevier.com/spatial-statistics}}}
%
%

\author[1]{Dionissios~T.~Hristopulos,
 \thanks{\texttt{dionisi@mred.tuc.gr.}; Corresponding author}}
\author[2]{Ivi~C.Tsantili\thanks{ \texttt{ivi.tsantili@gmail.com}}}

\affil[1]{Geostatistics Laboratory, School of Mineral Resources Engineering\\ Technical University of Crete, Chania, 73100 Greece}
\affil[2]{Beijing Computational Science Research Center, Beijing, 100193 China}
\maketitle \vspace{-14pt}
\begin{abstract}
The generation of non-separable, physically motivated covariance functions is a theme of ongoing research interest, given that only a few classes of such functions are available.  We construct a non-separable space-time covariance function based on a diffusive Langevin equation. We employ ideas from statistical mechanics to express the response of an equilibrium (i.e., time independent) random field to a driving noise process by means of a linear, diffusive relaxation mechanism. The equilibrium field is assumed to follow an exponential joint probability density which is determined by a spatial local interaction model.  We then use linear response theory to express the temporal evolution of the random field around the equilibrium state in terms of a Langevin equation. The latter yields an equation of motion for the space-time covariance function, which can be solved explicitly at certain limits. We use the explicit covariance model obtained in one spatial dimension and time. By means of the turning bands transform, we derive a non-separable space-time covariance function in three space dimensions and time. We investigate the mathematical properties of this space-time covariance function, and we use it to model a dataset of daily ozone concentration values from the conterminous USA.
\end{abstract}


\newpage



\section{Introduction}
\label{sec:intro}

The formulation of mathematically valid and physically motivated covariance functions is an ongoing pursuit
in spatiotemporal statistics~\cite{Christakos92,Christakos98,Gneiting02,Cressie11}.  This is driven by the central role
of covariance functions in the estimation and simulation of space-time processes.
In particular, there is
strong interest in formulating non-separable space-time covariance models. This
research is motivated first by the inadequacy of separable covariance models to capture the
patterns in realistic space-time data~\cite{Christakos98,Stein05}. Secondly, space-time random fields often represent
the evolution of physical variables under a respective partial differential equation (PDE)
with stochastic components
(e.g. initial or boundary conditions,  coefficients, or  driving noise)~\cite{Whittle54,Heine55,Yaglom87,Jones97}.
Even in simple cases, this evolution leads to the development of non-separable space-time covariance functions.
A case in point is the one-dimensional diffusion of an initially parabolic concentration field in a medium with
uniform but random diffusion coefficient~\cite{Christakos98}.
This problem illustrates how the non-separability of the covariance function
emerges from the superposition of separable space-time eigenfunctions.

Since  the explicit solution of PDEs is in many cases not feasible,
non-separable covariances have also been constructed  by employing mathematical permissibility criteria,
thus circumventing the problem of solving an underlying PDE, e.g.~\cite{Gneiting02,deIaco02,Ma03,Kolovos04}.
Another possibility for constructing space-time covariance functions is by integrating
permissible spectral densities. However, the explicit integration of the spectral density is not
feasible in many cases~\cite{Stein05}. Recent reviews of spatiotemporal covariance functions are given
in~\cite{Kolovos04,Ma08,Cressie11}.

As argued above, space-time covariance functions can be obtained as solutions to PDEs.
These PDEs are associated with respective stochastic partial differential equations (SPDEs),
also known as Langevin equations, that describe the evolution of the
corresponding space-time random field~\cite{dth15b}. This perspective was
used by Whittle to derive the spatial Whittle-Mat\'{e}rn covariance functions~\cite{Whittle54}.
Another example is the space-time covariance function developed by Heine
as a solution of a parabolic SPDE with one time and one space dimension~\cite{Heine55}.
The connection between rational spectral densities and Langevin equations that represent the
corresponding random field is also discussed by Yaglom~\cite{Yaglom57}.
The same perspective was also pursued by engineers to develop methods
for the control of systems with random parameters~\cite{Spanos03}.

Various approaches for the construction of covariance functions are possible in a framework
that involves partial differential equations. These equations may pertain to the evolution of the random field,
in which case they constitute stochastic partial differential equations, or they may describe the
dynamics of the covariance function, in which case they are deterministic partial differential equations.

The \textit{SPDE approach} aims to efficiently generate random fields
which obey  the Whittle-Mat\'{e}rn covariance. In the SPDE method, the random field that satisfies the  Langevin
equation associated with the Whittle-Mat\'{e}rn covariance family is solved using a projection to a finite-dimensional function base~\cite{Rue11}.
The SPDE method can also be used to generate non-stationary random fields~\cite{Lindgren15}.
To our knowledge this approach has not been extended to explicitly include random fields with
space-time covariance functions.

The \textit{covariance PDE} approach
focuses on deriving space-time covariance functions by solving suitable PDEs.
If the SPDE equation that determines the evolution of the studied process driven by noise is known,
it can be used to derive an associated covariance PDE which represents the equation of motion (EOM) of
the covariance function. If the initial SPDE is linear, so is the associated PDE for the covariance
function. For stationary systems the covariance PDE is defined over the domain ${\mathcal D} \in \Rd \times \R$
of spatial and temporal lags. Hence, the
derivatives are taken with respect to the spatial and temporal lags instead of the position and time.
This approach has been used in statistical physics to study dynamic critical phenomena in
ideal systems where the structure of the SPDE is assumed to be known~\cite{Hohenberg79}.

Often, the Langevin equations (SPDEs) that determine the dynamic evolution of the studied process are not known.
In such cases,  the equation of motion  of the covariance
can not be derived from first principles. In other cases,
the respective covariance PDEs may be known but unsolvable by analytical means.
Then,  one feasible solution is to use a flexible enough \textit{surrogate PDE}  to model space-time correlations.
This  approach was followed in the development of Spartan spatial random fields~\cite{dth03,dthsel07} and their space-time extensions~\cite{dth15b}.

The physical justification of surrogate PDEs is that they can provide tractable, albeit idealized, models for space-time correlations
which approximate the behavior of  broad classes of stochastic systems. For example, Heine~\cite{Heine55} considered
SPDEs of  elliptic, parabolic, and hyperbolic types which conform with the standard classification of partial differential equations with second-order partial derivatives~\cite{Dennis15}.
The types above are respectively associated with equilibrium (steady-state), diffusive, and wave propagation processes.
Hence, solutions for covariance functions that correspond  to SPDEs of these types can provide useful approximations for
general processes with the above characteristics. In a sense, the idea of using surrogate models is not new in spatial statistics:
all the classical covariance functions represent plausible but simplified forms of spatial dependence which can be used to
model, even if approximately, various patterns of spatial (or spatiotemporal) dependence.

 In simple cases the covariance PDEs
(either exact or surrogate) are linear and can be solved explicitly, at least for certain
choices of initial and boundary conditions. Explicit solution of covariance PDEs
was successfully employed  in the Whittle-Mat\'{e}rn
case~\cite{Whittle54}, in the Heine covariance model~\cite{Heine55}, and in the case of Spartan spatial random fields~\cite{dthsel07,dth15}.
Explicit results  for space-time covariance
functions were also recently derived using the tools of linear response theory~\cite{dth15b}.
The linear response theory and the Heine approach involve different theoretical frameworks. However,
in one spatial dimension the linear response theory
of the Spartan spatial random field recovers at the zero-curvature limit the parabolic Heine covariance model.

If the covariance PDE (exact or surrogate) is not amenable to explicit solution,
 approximate space-time covariance functions can be obtained by
applying model order reduction techniques.
Covariance PDEs based on model order reduction  simplify the mathematical problems using
Galerkin projections of the solution  on finite-dimensional bases~\cite{Dennis15}.
Order reduction methods have been used in engineering mechanics to
develop expansions of random fields when the dynamic equations satisfied by the field are known~\cite{Spanos03,Xiu03}.
Order reduction methods are broadly based on the same idea as the SPDE approach.
Once the reduced-order random field representation
is known, approximations of the covariance function can also be derived using
the most important basis functions.

Herein we construct a space-time covariance function based on the surrogate PDE approach.
The PDE is obtained within the
statistical mechanics framework of linear response theory. Our starting point is an equilibrium
Spartan spatial random field (SSRF)~\cite{dth03b}. This leads to a parabolic PDE for the covariance function, which is
suitable for diffusive processes. The remainder of this document is structured as follows: Section~\ref{sec:lrt}
presents an overview of the linear response theory leading to an equation of motion for the space-time covariance
function. Section~\ref{sec:tb} presents the solution of the above equation in $1+1$ dimensions, its transformation
using the turning bands method to a $3+1$ non-separable covariance function, and an investigation of its properties.
This section includes the new space-time covariance function~\eqref{eq:new-st} which is the main
closed-form result obtained in this work.
Section~\ref{sec:stslr-vario} develops the corresponding variogram function and obtains the respective space and time marginal variogram
functions.
Section~\ref{sec:estimation} focuses on the estimation of the new covariance parameters from space-time data
using the method of marginal space and time variograms.
These are used to  model the distribution of daily ozone values in the USA over a period of fourteen days in
Section~\ref{sec:data}.
 Finally, we present our conclusions and a  discussion of the results in Section~\ref{sec:conclu}.

\section{Review of Linear Response Theory}
\label{sec:lrt}

This section briefly reviews the main steps of the application of linear response theory in the construction of space-covariance functions
following~\cite{dth15b}.

Let us denote by $X(\bfs,t)$ the space-time random field at the space time point $(\bfs,t) \in \R^{d} \times \R$,
 and by $X(\bfs)$ the static field
 at equilibrium (in the absence of temporal fluctuations). Realizations of the field will be denoted by the lowercase letter
 $x(\bfs)$. The expectation over the ensemble of states of the field will be denoted by means of $\EE{\cdot}.$
 The spatial domain of interest is $\Do_{s} \subset \Rd$ and the temporal domain of interest $\Do_{T} \subset \R$.

\subsection{The equilibrium state}
 The equilibrium (static) regime provides an initial
condition on the space-time covariance function.
Since we are interested in covariance functions, we will focus on zero-mean random fields.
 First, we assume that the
field realizations in equilibrium are determined by the Gibbs probability density function (pdf)
with the following exponential form

\[
f_{\rm X}(\bfx;\bmthe) = {Z^{-1}(\bmthe)}\, \E^{- \Ha(\bfx;\bmthe)},
\]
\noi where $\Ha(\bfx;\bmthe)$ is a quadratic, non-negative functional of the field states (realizations) $\bfx = (x_{1}, \ldots, x_{N})^\top$
at $N$ points  $\bfs_{i} \in \Rd$,  where $i=1, \ldots, N$, the superscript $T$ denotes the transpose,
$\bmthe$ is the parameter vector, and
$Z(\bmthe)$ is the partition function which normalizes the joint density.

In the continuum limit $\bfx$ is replaced by the field $x(\bfs)$ which is defined at every site $\bfs$
on the spatial domain $\Do_{s}$.
The spatial correlations of the field realization $x(\bfs)$ are implemented by means
of the surrogate functional $\Ha[x(\bfs);\bmthe]$ which depends on
the values of the realization $x(\bfs)$ for all $\bfs \in \Do$.
For the energy functional we use the SSRF form~\cite{dth03}
with an additional parameter $\mu$ as follows

\beq
\label{eq:H-ssrf}
\Ha[x(\bfs);\bmthe] = \frac{1}{{2\eta_0 \xi ^d }} \int_{\Rd} d\bfs \, \left\{
\left[x(\bfs)\right]^2  + \eta_1 \,\xi^2
\left[ {\nabla x(\bfs)} \right]^2
 + \mu\, \xi^4 \, \left[ {\nabla^2 \, x(\bfs)} \right]^2 \right\},
\eeq

\noi where $\bmthe = (\eo, \e, \xi, \mu)^\top$.
In the energy function~\eqref{eq:H-ssrf} the first term in the integrand measures the square of the fluctuations, the second term is proportional to
the square gradient, and the third term to the square of the curvature (we assume that the curvature is represented by the Laplacian).
The coefficients in~\eqref{eq:H-ssrf} are selected so that the ``energy'' $\Ha[x(\bfs);\bmthe]$ is dimensionless.
Hence, the \textit{characteristic length} $\xi$ has units of length,
the \textit{rigidity} $\e$ is a dimensionless number which determines the resistance of $x(\bfs)$ to
bending, while $\eo$ has the units  $[X]^2$. Finally, the new dimensionless
parameter $\mu \ge 0$ controls the contribution of the curvature term.
 We refrain from calling  $\xi$ the correlation  length,
because the presence of $\e$ implies that the relation of $\xi$ with the correlation length is
more complex than for simpler models~\cite{serra11}.

The permissibility conditions imposed by Bochner's theorem~\cite{Bochner59} constraints for the values of $\e$. For
$\mu=1$ the permissibility condition is  $\e>-2$. For $\mu=0$, however, the respective condition
is more stringent since it requires $\e>0$. We will henceforth assume  $\e>0$ to account for all possible values of $\e$.

\subsection{The linear response Langevin equation}
The premise of
linear response theory is that the dynamic evolution of the system near the equilibrium state is
essentially determined by $\Ha[x(\bfs);\bmthe]$. In particular, if noise tends to drive the
system away from the equilibrium, the system will develop in response a restoring velocity that
depends on the departure from the equilibrium. The response is described in terms of the
following Langevin equation
\begin{equation}
\label{eq:noneq}
\frac{\partial x(\bfs,t)}{\partial t}= - \Gamma \,
\left. \frac{\delta \Ha[x(\bfs);\bmthe]}{\delta x(\bfs)} \right|_{x(\bfs)= x(\bfs,t)} + \zeta(\st),
\end{equation}
\noi where $\Gamma>0$ is a diffusion coefficient that determines the relaxation towards the equilibrium,
$\de(\cdot)/\de x(\bfs)$ is the functional derivative with respect to the field realization,
and $\zeta(\bfs,t)$ is  the
noise field. For the latter, we assume that it is a  Gaussian
white noise with $\EE{\zeta(\st)}=0$ and  $\EE{\zeta(\st)\, \zeta(\stp)} = D$, where
$D>0$ is the variance, and $\EE{\cdot}$ denotes the expectation with respect to the noise.

Equation~\eqref{eq:noneq} essentially determines the rate at which the random field realizations
change in time
as a superposition of two terms: the first component is the relaxation component which tends to restore the
 equilibrium, while the second component is a  stochastic velocity that perturbs the approach
to equilibrium.


\subsection{Equation of motion of the linear response covariance function}
A number of technical steps described in~\cite{dth15b} lead to the following partial differential equation
that determines the  motion of the
covariance function
\begin{equation}
\label{eq:cor-fun1}
\frac{\partial C(\bfs-\bfs', t-t'; \bmthe)}{\partial \tau} = - \Gamma \,\sign(\tau) \,
\EE{ x(\bfs,t) \, \left. \frac{\delta \Ha[x(\bfs);\bmthe]}{\delta x(\bfs')}\right|_{x(\bfs')=x(\bfs',t')}  }.
\end{equation}
In the above, $\tau = t -t' \in T$ is the temporal lag.  The sign function is defined by $\sign(\tau)=1$ if $\tau>0$,
$\sign(\tau)=-1$ if $\tau<0$, and $\sign(\tau)=0$ if $\tau=0$. The subscript $x(\bfz)=x(\bfs',t')$ denotes that
after the functional derivative is calculated, the static realization $x(\bfs')$ is replaced by the dynamic realization
$x(\bfs',t')$. The vector $\bmthe$ includes a general set of parameters; the parameter $\Gamma$ can be absorbed
in the amplitude coefficient of $\Ha[x(\bfs);\bmthe]$, i.e., in $\eo$.

Next, we replace  $\Ha[x(\bfs);\bmthe]$  in equation~\eqref{eq:cor-fun1} with the Spartan energy functional~\eqref{eq:H-ssrf},
evaluate the functional derivatives, and calculate the resulting field expectations (see~\cite{dth15b}). Then,
the covariance equation of motion is given by  the following linear PDE which includes fourth-order spatial derivatives
\begin{equation}
\label{eq:cor-eom}
\frac{\partial \cors}{\partial \tau} = - \frac{\sign(\tau)}{\tau_{c}} \,
\left(  1 - \eta_{1} \xi^{2}  \nabla^{2} + \mu\,\xi^{4}
\nabla^{4} \right) \, \cors,
\end{equation}
with suitable boundary and initial conditions which are defined below. We will refer to it as the
Spartan linear response PDE.

The function
 $\cors: \,\R^d \times T \to \R$ is the space-time covariance function associated with the linear response of the
 Spartan energy functional,  $\bfr = \bfs - \bfs' \in \R^d$
 denotes the spatial lag, and $\bmthe' = (\eo, \e, \xi, \mu, \tau_c)$ is the parameter vector of the covariance function.
 The \textit{characteristic time constant} $\tau_c$ is linked to the noise variance and the SSRF parameters by
 means of $\tau_c = D/2\eo \xi^d$.
 In addition, $\nabla^{2}=\sum_{i=1}^{d}\partial^{2} \big/ \partial r_{i}^{2}$ denotes the Laplacian operator,
 and its square $\nabla^{4} = (\nabla^{2})^{2}$ is the biharmonic operator~\cite{Dennis15}.
The dependence on $\eo$ is introduced via the initial condition as shown below.

In the Spartan linear response PDE~\eqref{eq:cor-eom} the
term proportional to $\nabla^{2}$ is derived via the functional differentiation
from the $\Ha[x(\bfs)]$ component that involves the
integral of the square gradient $ [\nabla x(\bfs)]^2$, whereas the term proportional to $\nabla^{4}$ is obtained from
the term proportional to the integral of the square curvature $ [\nabla^{2} x(\bfs)]^2$ of the field over
$\Do_s$.

\subsection{Solution of the Spartan linear response PDE based on the spectral method}

Assuming that $\Do_s$ expands to an infinite support, the \textit{boundary conditions}  are
that $\cors$ tends to zero as $\rr \to \infty$ (unbounded domain).
Then, the Spartan linear response PDE~\eqref{eq:cor-eom} is best solved by applying
 the Fourier integral transform method~\cite{King03} to the spatial component of
the covariance function. The \textit{spatial Fourier transform} of the covariance function is given by

\[
\tilde{C}(\bfk,\tau; \bmthe') = \mathcal{F}_{\bfr}[C(\bfr,\tau; \bmthe')] = \int_{\Rd} d\bfr  \, \E^{-\jmath \bfk \cdot \bfr} \, C(\bfr,\tau; \bmthe'),
\]
where  the function
$\tilde{C}(\bfk,\tau;\bmthe')$ is the time-dependent spectral density
where $\bfk \in \Rd$ is the wavevector  in reciprocal space.
The \textit{inverse Fourier transform (IFT)} is then given by

\[
C(\bfr,\tau;\bmthe') = \mathcal{F}^{-1}_{\bfk}[\tilde{C}(\bfk,\tau;\bmthe')]= \frac{1}{(2\pi)^d} \,
\int_{\Rd} d\bfk  \, \E^{\jmath \bfk \cdot \bfr} \, \tilde{C}(\bfk,\tau;\bmthe').
\]

Inserting the IFT in the Spartan linear response  PDE~\eqref{eq:cor-eom} leads to the following
\textit{first-order ordinary
 differential equation (ODE)} with respect to time with initial condition $\tilde{C}(\bfk,0)$

 \begin{subequations}
 \label{eq:ode}
 \beq
\label{eq:eom-f-fgc-cor}
\frac{\partial \tilde{C}(\bfk,\tau;\bmthe')}{\partial \tau}= -  \frac{\sign(\tau)}{\tau_c} \,
\left(  1 + \eta_{1} k^2\xi^{2}  + \mu\, k^4\xi^{4}\right) \, \tilde{C}(\bfk,\tau;\bmthe').
\eeq

The above ODE is solved using  the statistically homogeneous and isotropic
\textit{initial condition}. This is provided by
 the Spartan spectral density which corresponds to the equilibrium random field~\cite{dth03}, i.e.,
 \beq
\label{eq:ssrf-spd}
\tilde{C}(\kk,0;\bmthe') :=  \tilde{G}(\bfk)= \frac{\eo\, \xi^d}{1 + \e ( \kk \xi)^2 + \mu  (\kk \xi)^4  },
\eeq
\end{subequations}
where $\kk \in \R_{+}$ is the wavenumber, i.e.,
the Euclidean norm of the reciprocal-space wavevector $\bfk$.

The solution of the linear ODE~\eqref{eq:eom-f-fgc-cor} for the time-dependent spectral density is  given by
the exponential function
\beq
\label{eq:eom-ft-fgc-cor}
\tilde{C}(\kk,\tau; \bmthe') =
\tilde{G}(\bfk) \, \E^{- \left(1 + \eta_{1} \kk^{2}
\xi^{2} + \mu\, \kk^{4} \xi^{4} \right) \, \vert\tau\vert /\tau_c }.
\eeq
Real-space solutions are obtained by evaluating the inverse Fourier transform of~\eqref{eq:eom-ft-fgc-cor}
using the isotropic spectral representation~\cite{Yaglom87}.

\subsubsection{The zero-curvature limit}
For $\mu=0$ (zero-curvature model), the covariance equation of motion~\eqref{eq:cor-eom}
is a \textit{second-order PDE of the parabolic type}. In~\cite{dth15b} we solved this PDE
using the Fourier integral transform method~\cite{King03}.
The expressions derived for the covariance function in
real space depend on the dimensionality $d$. For $d=2,3$ the derived functions have singularities
at the origin.
In contrast, the covariance function is well defined in $d=1$.

\subsection{Properties of solutions of the Spartan linear response PDE}
Properties of the functions $\cors$ that solve the Spartan linear response PDE~\eqref{eq:cor-eom}
can be investigated,
even if explicit forms of the solutions are not available.

\begin{enumerate}

\item  Space-time covariance functions generated by the linear response of the Spartan energy functional are
 statistically homogeneous in space and stationary in time. These properties are due to the
choices of  infinite support, constant coefficients, and homogeneous initial condition.

\item In addition, the solution is $\cors$ isotropic, because neither the Spartan linear response PDE~\eqref{eq:cor-eom} nor the
initial condition~\eqref{eq:ssrf-spd} distinguish between different spatial directions.
This constraint can in principle be easily relaxed by replacing $\kk \, \xi$ in the
ODE and the initial condition~\eqref{eq:ode} with $\sum_{i=1}^{d} k_{i} \, \xi_{i}$.
In the anisotropic case, however, the analytical evaluation of the inverse Fourier transform has not
been done.

\item A bounded domain
with non-periodic boundary conditions will lead to non-homogeneous
covariance functions. Explicit solutions would be more difficult to obtain in such cases.

 \end{enumerate}

\section{Space-Time Covariance Functions by means of the Turning Bands Method}
\label{sec:tb}

In this section, we apply the turning bands method to the one dimensional (in space)
covariance function~\eqref{eq:covariace-d-1-m-0}. This operation  generates a permissible space-time covariance function in
three spatial dimensions and time.

 \subsection{Space-time covariance function in $1+1$ dimensions}

The one-dimensional space-time covariance function $C_{1}(r,\tau)$
based on the surrogate Spartan
functional $\Ha[x(\bfs);\bmthe]$ defined by~\eqref{eq:H-ssrf} is given by the following expression

\begin{align}
\label{eq:covariace-d-1-m-0}
C_{1}(h,u;\tilde{\bmthe}) & = \frac{\eta_{0}\, \la}{4}\,\left[ \E^{-\la\, h}
\erfc\left(\sqrt{u}-\frac{\la\, h}{2\sqrt{u}}\right) + \, \E^{\la\, h}
\erfc\left(\sqrt{u}+\frac{\la\, h}{2\sqrt{u}}\right)\right].
\end{align}

In the above, $h= |r|/\xi$ and $u=|\tau|/\tau_c$ are the normalized space and time lags and
$\tilde{\bmthe}=(\eo, \la, \xi, \tau_c)^\top$ is the parameter vector.
The \textit{flexibility (inverse rigidity)} constant is
$\la=1/\sqrt{\e}$, and $\erfc(\cdot)$ is the complementary error
function defined by the following integral
\[
\erfc(x) = \frac{2}{\sqrt{\pi}}\int_{x}^{\infty} dt \,\E^{-t^2}.
\]
The equation~\eqref{eq:covariace-d-1-m-0}  recaptures the covariance model of Heine~\cite{Heine55,Jones97},
albeit  with a different parametrization.

\subsection{Turning bands method}

Space transforms are mathematical operations that can
generate higher-dimensional functions based on lower-dimensional projections~\cite{Ehrenpreis03}.
In particular, Matheron developed the \textit{turning bands method} which can be
used to produce higher-dimensional isotropic covariance and generalized covariance functions from
one-dimensional covariances~\cite{Matheron73}.

We  briefly describe the turning bands method following the presentation in~\cite{Matheron73}.
Let  $Y(t)$, where $t \in \R$ be
a one-dimensional random process with covariance function $C_{1}(t-t')$ and let
$\mathbf{p} \in \Rd$ be a unit random vector with $\|\mathbf{p}\|=1$.
Moreover, let an isotropic random field $X(\bfs)$, where $\bfs \in \Rd$ be such that
$Y(t) = X_{\mathbf{p}}(\bfs)=X(\bfs \cdot \mathbf{p})$ is the one-dimensional projection of
$X(\bfs)$ along the vector $\mathbf{p}$. The covariance function of the projection of
the random field $X(\bfs)$ is given by
\[
\EE{X_{\mathbf{p}}(\bfs) \, X_{\mathbf{p}}(\bfs')} = C_{1}\left(\mathbf{p} \cdot (\bfs - \bfs')\right).
\]
The covariance of $X(\bfs)$ can be evaluated by calculating the average over one-dimensional projections
along all different directions $\mathbf{p}$, i.e.,

\[
C(\rr) = \int_{B_d} \D\mathbf{p} \, C_{1}\left(\mathbf{p} \cdot \bfr \right) \, \pi(\mathbf{p}),
\]
where $\pi(\mathbf{p})$ is the probability density function representing the distribution
of the unit vector $\mathbf{p}$, and
$B_d$ denotes that the space integral is evaluated
over the surface of the unit sphere. Then it follows that

\beq
\label{eq:tb}
C_{d}(\rr) = \frac{2\, \Gamma(d/2)\, \pi^{-1/2}}{\Gamma\left(\frac{1}{2} (d-1) \right)}
\, \int_{0}^{1} C_{1}(v \rr)\, (1 - v^2)^{(d-3)/2}.
\eeq

\subsection{Non-separable space-time covariance function}
In light of the above, and in particular equation~\eqref{eq:tb},
a \textit{three-dimensional, isotropic covariance} function is generated from $C_{1}(r,\tau)$
by means of the following integral (see also~\cite{Mantoglou82})
\[
C_{3}(\rr,\tau;\tilde{\bmthe}) = \frac{1}{\rr} \int_{0}^{\rr} \D x \, C_{1}(x, \tau;\tilde{\bmthe})= \frac{1}{h} \int_{0}^{h} \D y \, C_{1}(y, u;\tilde{\bmthe}),
\]

\noindent where $C_{1}(\cdot,\cdot;\tilde{\bmthe})$ is given by~\eqref{eq:covariace-d-1-m-0}. The above integral
requires the integration the $\erfc(x)$ function, which we perform  using integral tables~\cite{Ng69}.
The integration leads to the following
\textit{non-separable} space-time covariance function

\begin{align}
\label{eq:new-st}
 C_{3}(h,u;\tilde{\bmthe}) = & \frac{\eta_{0}}{4 h}
\left[  2\, \E^{-u}\, \erf\left( \frac{\la h}{2\sqrt{u}} \right) +  \E^{\la\, h}\, \erfc\left( \sqrt{u} + \frac{\la h}{2\sqrt{u}}\right)  \right.
\nonumber \\
    &  \quad  \left. - \E^{-\la\, h}\,
\erfc\left( \sqrt{u} -\frac{\la h}{2\sqrt{u}}\right) \right],
 \end{align}
\\
where $h=\| \bfr \|/\xi$ is the normalized spatial lag, $u = |\tau|/\tau_c,$ is the normalized time lag, $\erfc(x)$ is the complementary error function,
and  $\erf(x)= 1- \erfc(x)$, is the error function.
The above is the space-transformed image of the $1+1$ zero-curvature, linear response covariance function~\eqref{eq:covariace-d-1-m-0}.
Hence, we will refer to it as the ``\textit{Space-Transformed  Spartan Linear Response}'' or STSLR covariance function.

 \subsection{Properties of the STSLR space-time covariance function}

In this section we investigate some properties of the space-transformed image of the $1+1$
zero-curvature, linear response covariance function. As we have already noted above, the STSLR
covariance function~\eqref{eq:new-st} is \textit{non-separable, stationary in time, and spatially isotropic.}
In the following, we drop the dependence of the covariance function on $\tilde{\bmthe}$ for notational brevity.

A general class of non-separable space-time covariance functions is provided by the
isotropic Gneiting family~\cite{Gneiting02},
\[
C(h,u) = \frac{\al^2}{\psi(u^2)^{\de}} \phi\left( \frac{h^2}{\psi(u^2)} \right),
\]
where $\phi(\cdot)$ is a completely monotone function,  $\psi(\cdot)$ is a  Bernstein function
(i.e., a positive  function with completely monotone derivative), $\al>0$ and $\de >d/2$.
Both $\phi(\cdot)$ and $\psi(\cdot)$ satisfy appropriate boundary conditions at zero and infinity.
We note that the STSLR covariance has a space-time dependence which is not included in the broad class of
Gneiting functions.

The STSLR model also differs from the extension of the Heine $\R \times T$ model to $\Rd \times T$ which was
proposed by Ma~\cite{Ma03,Stein05}; in addition to the different arguments of the complementary error function,
the STSLR model includes a third term which involves the error function.

\subsubsection{Symmetry}
\begin{figure}[htbp]
	\begin{center}
		\includegraphics[width=0.8\textwidth]{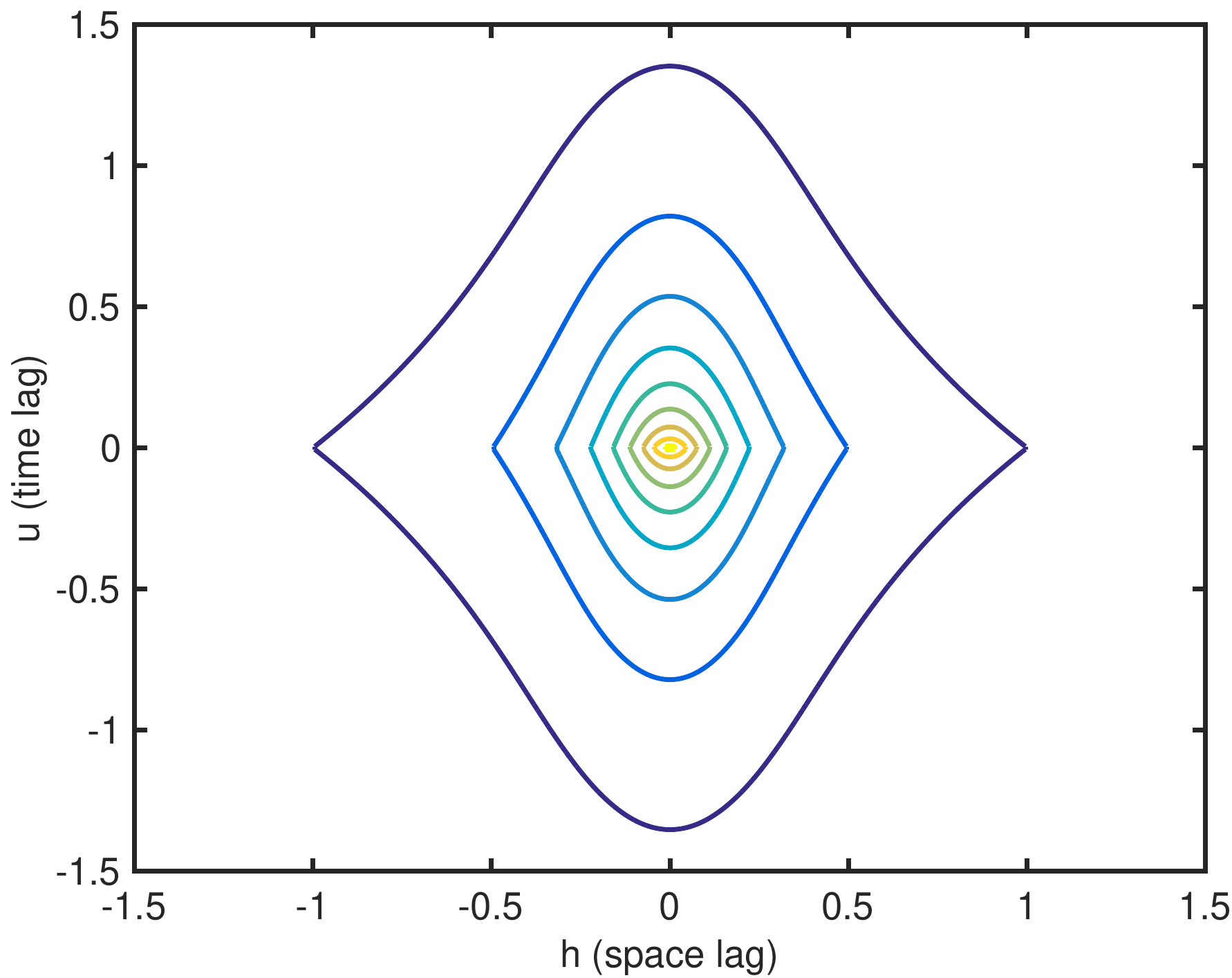}
		\caption{Contour plots of the STSLR space-time covariance function~\eqref{eq:new-st} generated by applying the turning bands method to the $1+1$ covariance function~\eqref{eq:covariace-d-1-m-0} which is obtained by the linear response of the zero-curvature SSRF energy function.}
		\label{fig:cov_st_sli}
	\end{center}
\end{figure}

The STSLR covariance function~\eqref{eq:new-st} depends on the spatial and temporal lags only through their magnitudes.
Hence, it satisfies the properties $C_{3}(\bfr,\tau)= C_{3}(-\bfr,\tau)$ and
 $C_{3}(\bfr,\tau)= C_{3}(\bfr,-\tau)$ and is a \textit{fully symmetric covariance function} according to~\cite{Gneiting02,li07}.
 The contours of $C_{3}(\bfr,\tau)$ in the domain $ \R \times T$  are shown in Fig.~\ref{fig:cov_st_sli}.

\begin{figure}[htbp]
	\begin{center}
		\includegraphics[width=0.9\textwidth]{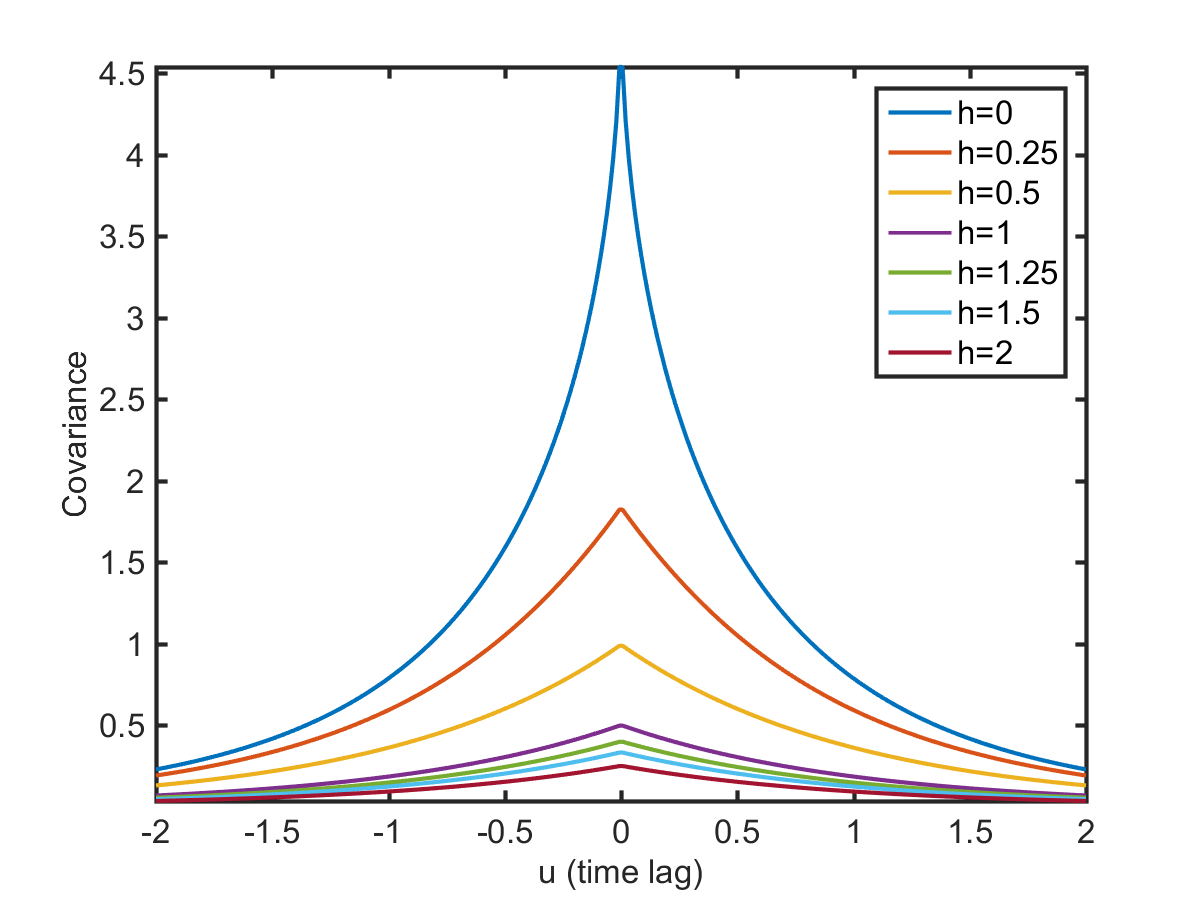} %
		\caption{Parametric plots of the STSLR covariance~\eqref{eq:new-st}
        versus the normalized time lag $u$ for different space lags $h$; the space lag increases in the direction from top to bottom.}
		\label{fig:cov_st_sli_time}
	\end{center}
\end{figure}

\begin{figure}[htbp]
	\begin{center}
		\includegraphics[width=0.9\textwidth]{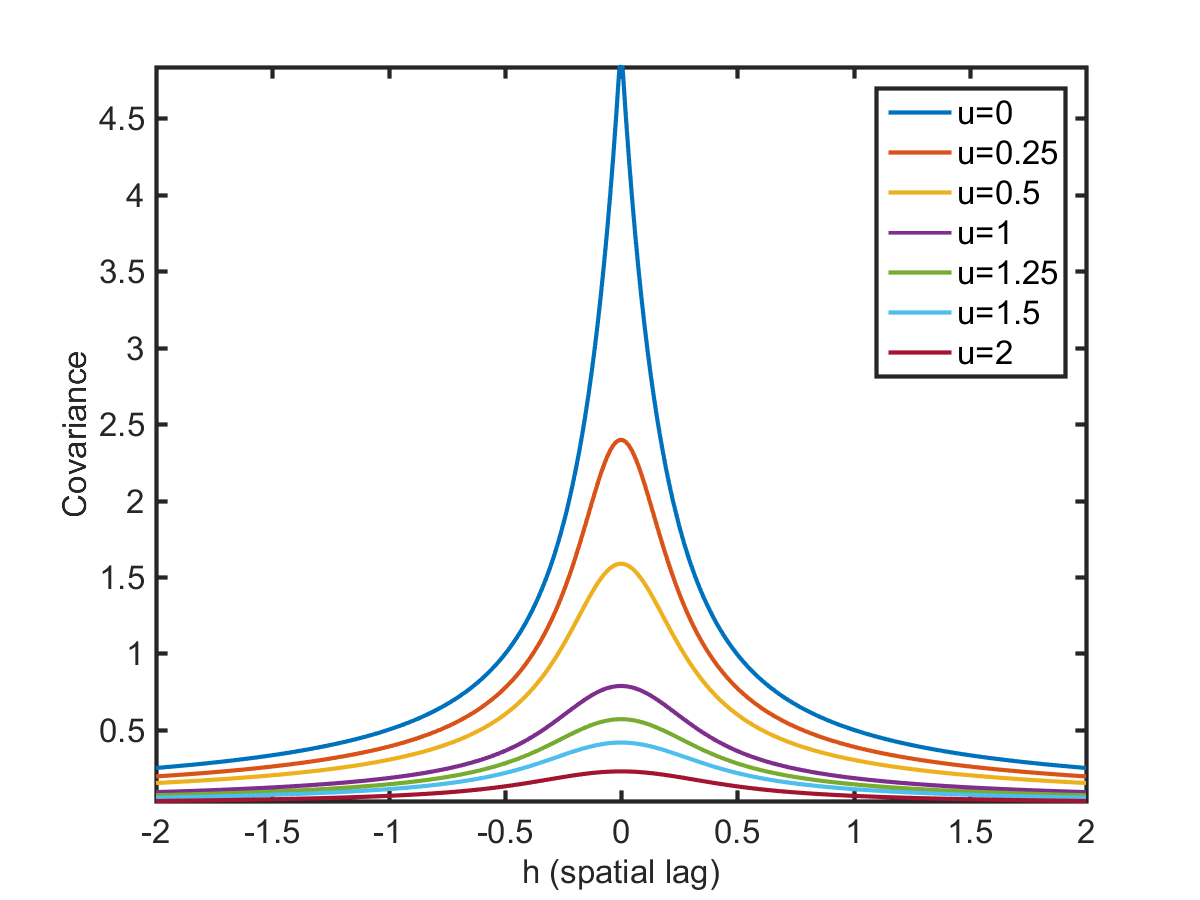} %
		\caption{Parametric plots of the STSLR covariance~\eqref{eq:new-st}
    versus the normalized spatial lag $h$ for different time lags $u$; the time lag increases in the direction from top to bottom.}
		\label{fig:cov_st_sli_spatial}
	\end{center}
\end{figure}

\subsubsection{Continuity and differentiability}
The STSLR covariance function~\eqref{eq:new-st} is continuous for every (normalized) space lag $h$ (time lag $u$).
We show this as follows: For all $u \in R_{0,+}$ and  $h \in R_{+}$,  the function
$C_{3}(h,u)$ is continuous because it comprises products of
everywhere continuous functions,
e.g., $\exp(\cdot)$, $\erf(\cdot)$, and $\erfc(\cdot)$.
The case $h=0$ should be treated separately, because of the factor $1/h$ which diverges as $h \to 0$.
The limit $ h \to 0$ is evaluated in~\eqref{eq:cov-time} below, leading to
the marginal covariance function
$C_{T}(u)$ which is continuous for all $u \in \R_{0,+}$.
Hence, the continuity of $C_{3}(h,u)$  is assured for all $h,u \in \R_{+,0}$.

In addition, each of the three summands in the covariance~\eqref{eq:new-st}
involves either (i) $\erf(\cdot)$ or $\erfc(\cdot)$ and (ii)
either $\exp(-u)/h$ or $\exp(\pm h)/h$. The functions (i)  are everywhere differentiable,
whereas the functions (ii) are non-differentiable at $h=0$ or both $u=0$ and $h=0$ due
the presence of the absolute values in $h$ and $u$.

The above continuity and differentiability properties imply that  random fields
with the covariance~\eqref{eq:new-st} are continuous but non-differentiable in the mean square sense~\cite{Adler10}.

 Parametric plots of the STSLR covariance function $C_{3}(h,u)$ for fixed spatial lags are shown in Fig.~\ref{fig:cov_st_sli_time} and
 for fixed time lags in Fig.~\ref{fig:cov_st_sli_spatial}. Note that  $C_{3}(h,u)$ for fixed $u \neq 0$
 appears smooth everywhere (including $h=0$), while for every fixed $h$ the  function $C_{3}(h,u)$ has a cusp at the origin
 $(u=0)$.

 This behavior is due to the fact that $C_{3}(h,u)$ admits a Taylor expansion around $h=0$ for finite $u$ but not
 around $u=0$ for finite $h$. The Taylor expansion around $h=0$ is given by
 \[
 C_{3}(h,u) = \frac{\eo \la}{2}\, \erfc\left(\sqrt{u}\right) + f_{1}(u)\,h^2 + f_{2}(u)\, h^4 + \Or(h^6),
 \]
where the functions $f_{i}(u)$, $i=1, 2, \ldots,$ diverge as $u\to 0$. However, for $u \neq 0$ the Taylor expansion is well defined
and only involves terms $\Or(h^{2n})$, where $n$ is an integer. This property implies that the field fluctuations are smoother in space
than in time~\cite{Stein05}.

A Taylor expansion is not possible around $u=0$ even for $h \neq 0$,
due to the presence of the factors $u^{-1/2}$ in the $\erf(\cdot)$ and $\erfc(\cdot)$ terms of
the covariance function~\eqref{eq:new-st}.

\section{The STSLR Space-Time Variogram Function}
\label{sec:stslr-vario}
In this section we derive the STSLR variogram that corresponds to the stationary covariance function~\eqref{eq:new-st} as well
as expressions for the marginal time and space covariance and variogram functions.

The STSLR covariance function~\eqref{eq:new-st} is reduced to simpler expressions at zero spatial and temporal
lag. These are useful in the estimation of the covariance parameters from space-time data.
The spatial marginal STSLR function $C_{S}(h)$ and the  temporal marginal STSLR function $C_{T}(u)$
are given by the following limits

\begin{subequations}
\beq
\label{eq:ct-def}
C_{T}(u) \doteq  \lim_{h \to 0}C_{3}(h,u).
\eeq
\beq
\label{eq:cs-def}
C_{S}(h) \doteq  \lim_{u \to 0} C_{3}(h,u),
\eeq
\end{subequations}

Based on these, we also define the respective marginal spatial and temporal STSLR variogram functions as follows
\beq
\gamma_{S}(h) = C_{S}(0) - C_{S}(h), \quad \gamma_{T}(h) = C_{T}(0) - C_{T}(h).
\eeq

\subsection{Temporal marginal functions}
To evaluate $C_{T}(u)$ based on the limit~\eqref{eq:ct-def} we use the Taylor series expansion of the error function around zero, i.e.,
\[
\erf(x) = \frac{2}{\sqrt{\pi}} \, \left( x - \frac{x^3}{3} + \Or(x^5) \right),
\]
where the notation $\Or(x^p)$ implies that the omitted terms are of order $p$ or higher.
Based on this expansion we evaluate the ratio of the error function over $h$ as follows
\[
\frac{1}{h} \erf\left( \frac{\la h}{2\sqrt{u}} \right) =  \frac{\la }{\sqrt{\pi\,u}} + \Or\left( h^{2}\right).
\]
The above leads to an expression that is  independent of $h$ at the limit $h \to 0$, i.e.,
\beq
\label{eq:time-lim-1}
\lim_{h \to 0} \frac{\E^{-u}}{ 2h} \erf\left( \frac{\la h}{2\sqrt{u}} \right) =  \frac{\la \, \E^{-u}}{2\sqrt{\pi\,u}}.
\eeq

Similarly, we  use the following Taylor expansion
for the complementary error function around $h=0$
\[
\erfc\left( \sqrt{u} \pm \frac{\la h}{2\sqrt{u}}\right) = \erfc\left( \sqrt{u} \right)  \mp \frac{\la h}{\sqrt{\pi \, u}}\, \E^{-u}\, + \Or(h^2).
\]

\noindent In light of the above,  the terms in the STSLR covariance~\eqref{eq:new-st} that involve the
complementary error function behave at the limit $h \to 0$ as follows

\begin{align}
\label{eq:time-lim-2}
&   \lim_{h \to 0}  \frac{1}{4h} \left[ \E^{\la\, h}\, \erfc\left( \sqrt{u} + \frac{\la h}{2\sqrt{u}}\right)  - \E^{-\la\, h}\,
\erfc\left( \sqrt{u} -\frac{\la h}{2\sqrt{u}}\right) \right] =
\nonumber \\
& \lim_{h \to 0}  \left[ \frac{\sinh (\la h)}{2h}\,\erfc\left( \sqrt{u}\right)   -
 \frac{\la}{2\sqrt{\pi \,u}}\, \E^{-u} \cosh (\la h) \right]
\nonumber \\
& =  \frac{\la}{2} \,  \erfc\left( \sqrt{u}\right) - \frac{\la}{2\sqrt{\pi \,u}}\, \E^{-u}  .
\end{align}

\noindent
By combining the expansions for the covariance terms that are proportional to the error
function and the complementary error function, i.e., equations~\eqref{eq:time-lim-1} and~\eqref{eq:time-lim-2}, we obtain
the following expressions for the temporal marginal covariance $C_{T}(u)$ and the temporal marginal variogram $\gamma_{T}(u)$

\begin{subequations}
\begin{align}
\label{eq:cov-time}
C_{T}(u) = \frac{\la\, \eta_{0}}{2} \,  \erfc\left( \sqrt{u}\right),
\\[ 1ex]
\label{eq:vari-time}
\gamma_{T}(u) = \frac{\la\, \eta_{0}}{2} \, \left[ 1-   \erfc\left( \sqrt{u}\right)   \right].
\end{align}
\end{subequations}

\noindent The dependence of $C_{T}(u)$ on the normalized time lag is illustrated in the
parametric plot shown in Fig.~\ref{fig:cov_st_sli_time} for $h=0$. The marginal covariance $C_{T}(u)$ tends asymptotically
to zero as $u \to 0$.

\subsection{Spatial marginal functions}
The marginal spatial covariance is obtained from~\eqref{eq:new-st}.
At zero temporal lag, i.e.,  $u \to 0$ and finite $h$  the arguments of the error function and the complementary error function
 in~\eqref{eq:new-st} tend to infinity. Thus, taking into account that
$\erfc(x)=1 - \erf(x)$ as well as the limits
$\lim_{x \to \infty} \erf(x) =1$, $\lim_{x \to \infty} \erfc(x) =0$ and $\lim_{x \to -\infty} \erfc(x) =2$,
 we obtain the following expressions for the marginal spatial covariance and variogram functions

\begin{subequations}
\begin{align}
\label{eq:cov-space}
C_{S}(h) = \frac{\eta_{0}}{2\, h} \left(  1 - \E^{-\la\, h}  \right),
\\[1ex]
\label{eq:vari-space}
\gamma_{S}(h) = \frac{\eta_{0}}{2}  \left[ \la - \frac{1 - \exp(-\la h)}{h}   \right].
\end{align}
\end{subequations}

\noindent The dependence of $C_{S}(h)$ on the normalized spatial lag is illustrated in the
parametric plot shown in Fig.~\ref{fig:cov_st_sli_spatial} for $u=0$.

\subsection{The STSLR Variogram Function}

The space and time marginal covariances tend both to the variance as the respective lag tends to zero,
i.e., $\lim_{h \to 0}C_{S}(h) = \lim_{u \to 0} C_{T}(u) =\sigma^{2}_{\rmx}$.
Hence, as it it follows from~\eqref{eq:cov-space} and~\eqref{eq:cov-time} the variance of the
spatiotemporal random field is $\sigma^{2}_{\rmx} = \eta_{0} \, \la/2$.
Consequently, the STSLR variogram function is given by
\beq
\label{eq:vario-sli}
\gamma(h,u;\tilde{\bmthe} )= \frac{\eta_{0}\la}{2} -  C_{3}(h,u;\tilde{\bmthe}),
\eeq
where $C_{3}(h,u;\tilde{\bmthe})$ is given by~\eqref{eq:new-st} and $\tilde{\bmthe}^\top = (\eta_0, \la, \xi, \tau_{c})$.

 \section{Estimation}
\label{sec:estimation}

In this section we discuss the estimation of the parameters $\tilde{\bmthe}$ of the STSLR variogram
from available space-time data.
In principle it is possible to use maximum likelihood to estimate the optimal parameters from the
data.  However, maximum likelihood is computationally intensive due to  memory storage requirements
which scale as $\Or(N^2)$, and the computational complexity of the covariance inversion which scales as
$\Or(N^3)$ for dense matrices, where $N$ is the sample size.
 There exist methods that  address the computational burden of maximum likelihood estimation by
 means of approximations or by dividing  the problem in smaller pieces~\cite{Stein04,Fuentes07}.
 Herein, we opt for a modified method of moments which is based on the marginal
 variograms and is simple to implement~\cite{Cesare01}.

We assume that there are  $N_{T}$ time instants and that $N_{S}$ is the
total number of spatial locations (stations) that report a measurement at least at one time instant.
The data involves the measurements $x_{i(j),j}$ where  $j=1, \ldots, N_{T}$ is the time index, and $i(j) \in \{1, \ldots N_{S} \}$
is the space index.  We   assume a fixed time step  $\de t$, so that $t_{j}= j\, \de t$, for $j=1, \ldots N_{T}$.
The normalized time lag then is $u_{k} = k\, \de t/\tau_c$, where $k=0, \ldots, N_{T}-1$.

For each time instant $t_{j}$, $j=1, \ldots N_{T}$, we assume that there exists at least one measurement, e.g., at the location $\bfs_{i(j)}$
where $i(j) \in \{1, \ldots, N_S \}$.
We denote by $S_{j}$ the set of points in space for which there are measurements at time $t_{j}$.
This set comprises the locations $S_{j}=\{ \bfs_{i(j)} \}$ where $i(j) \in \{1, \ldots, N_S \}$.
The cardinal number of the set is $N_{j} = \# S_{j}$. Since not all the stations have data at all times,
it holds that $N_{j} \le N_{S}$.  Let us also denote by $S_{j,m} = S_{j} \cap S_{m}$
the set  of spatial locations with measurements at both times $t_{j}$ and $t_{m}$, and by
$N_{j,m} = \#(S_{j} \cap S_{m})$ the cardinal number of this set.

We  estimate the empirical \textit{temporal marginal
variogram} as follows

\begin{subequations}
\label{eq:temporal-vario-estim}
\begin{align}
\hat{\gam}_{t}(t_{m},t_{m+k}) & =  \frac{1}{2 N_{m+k,m}} \, \sum_{\bfs_{i(m)} \in S_m} \,  \mathbb{I}_{S_{j,m}} \, \de_{j,m+k} \, \left[    x_{i(m),j} -  x_{i(m),m} \right]^2,
\\
& \mbox{for} \; m=1, \ldots, N_{T} -k, \; k=1, \ldots, \lfloor p\, N_{T} \rfloor, \nonumber
\\
\hat{\gam}_{T}(u_{k}) & =   \frac{1}{N_{S} - k} \,\sum_{m=1}^{N_{S}-k} \hat{\gam}_{t}(t_{m},t_{m+k}).
\end{align}
\end{subequations}
The indicator function $\mathbb{I}_{S_{j,m}}$ takes values $\mathbb{I}_{S_{j,m}}=1$ if both
$\bfs_{m}$ and $\bfs_{j}$ belong to the set $S_{j,m}$ and $\mathbb{I}_{S_{j,m}}=0$ otherwise. The
Kronecker delta, $\de_{i,j}=1$ if $i=j$ and $\de_{i,j}=0$ if $i \neq j$, restricts the respective
average to times $t_{j}$ and $t_{m}$ such that $j=m+k$.
Hence, the function $\hat{\gam}_{t}(t_{m},t_{m+k})$ is a purely spatial average over different locations for
two specific times which are $k$ steps apart.
The maximum lag is defined as a fraction $0 \le p \le 1$ of $N_{T}$, where $\lfloor p\, N_{T} \rfloor$ denotes the
largest integer that does not exceed $p\, N_{T}$.
 The function $\hat{\gam}_{T}(t_{m},t_{m+k})$ is a temporal average
of $\hat{\gam}_{t}(t_{m},t_{m+k})$ over all pairs of time instants that are $k$ steps apart.

Similarly, we define the estimator of the \textit{spatial marginal variogram} as follows

\begin{subequations}
\label{eq:spatial-vario-estim}
\begin{align}
\hat{\gam}_{s}(\bfs_{\al},\bfs_{\be}) & = \frac{1}{2N_{\al,\be}}\, \sum_{j=1}^{N_{S}}\, \mathbb{I}_{j;{\al,\be}} \,
 \left(    x_{\al,j} -  x_{\be,j} \right)^2, \; \al,\be = 1, \ldots, N_{S}
 \\
 N_{\al,\be} & = \sum_{j=1}^{N_{T}} \mathbb{I}_{j;{\al,\be}},
 \\
\hat{\gam}_{S}(\bfr) & = \frac{1}{N_{s}(\bfr)} \, \sum_{\al=1}^{N_{S}} \, \sum_{\be=1}^{N_{S}}
           \hat{\gam}_{s}(\bfs_{\al},\bfs_{\be}) \, \mathbb{I}_{\bfs_{\al},\bfs_{\be} \in B_{\epsilon}(\rr)}.
\end{align}
\end{subequations}
In the above,  $\mathbb{I}_{j;{\al,\be}}$ is an indicator function such that $\mathbb{I}_{j;{\al,\be}}=1$
if both locations $\bfs_{\al}$ and $\bfs_{\be}$ have measurements at the time instant $t_{j}$
and $\mathbb{I}_{j;{\al,\be}}=0$ otherwise.
$\mathbb{I}_{\bfs_{\al},\bfs_{\be} \in B_{\epsilon}(\rr)}$ is an indicator function that equals
one if the endpoint of the lag vector $\bfs_{\al}-\bfs_{\be}$ is inside
the neighborhood of the vector $\bfr$, which is defined in terms of a specified tolerance
 $\epsilon$.

By fitting $\hat{\gam}_{T}(u_{k})$ to the STSLR marginal temporal variogram~\eqref{eq:vari-time}
we estimate the  characteristic time $\tau_c$ and the
product $\la \, \eta_0$. Then, by fitting $\hat{\gam}_{S}(\bfr) $ to the marginal spatial STSLR variogram~\eqref{eq:vari-space}
we estimate  $\la$ and the spatial characteristic length $\xi$. Finally, using the estimate of the product
$\la \eta_{0}$ from the fit of the marginal temporal STSLR variogram, we estimate $\eta_0$.

%

\section{Application to Ozone Data}
\label{sec:data}
We use a set of daily concentration values of atmospheric ozone over the conterminous USA (excluding Alaska and Puerto Rico), which were
downloaded from the US Environmental Protection Agency Air Data website~\cite{epa}.
The data involve daily averages of ozone levels at 629 stations sampled over 14 consecutive days starting on January 1,
2015 (daily summary data file 44201 2015). The daily averages are based on eight measurements per day. The
ozone concentration is measured in parts per million (ppm) in volume. For the purpose of the analysis below the ozone concentrations
have been multiplied by 100.

\begin{figure}[htbp]
	\begin{center}
		\includegraphics[width=0.8\textwidth]{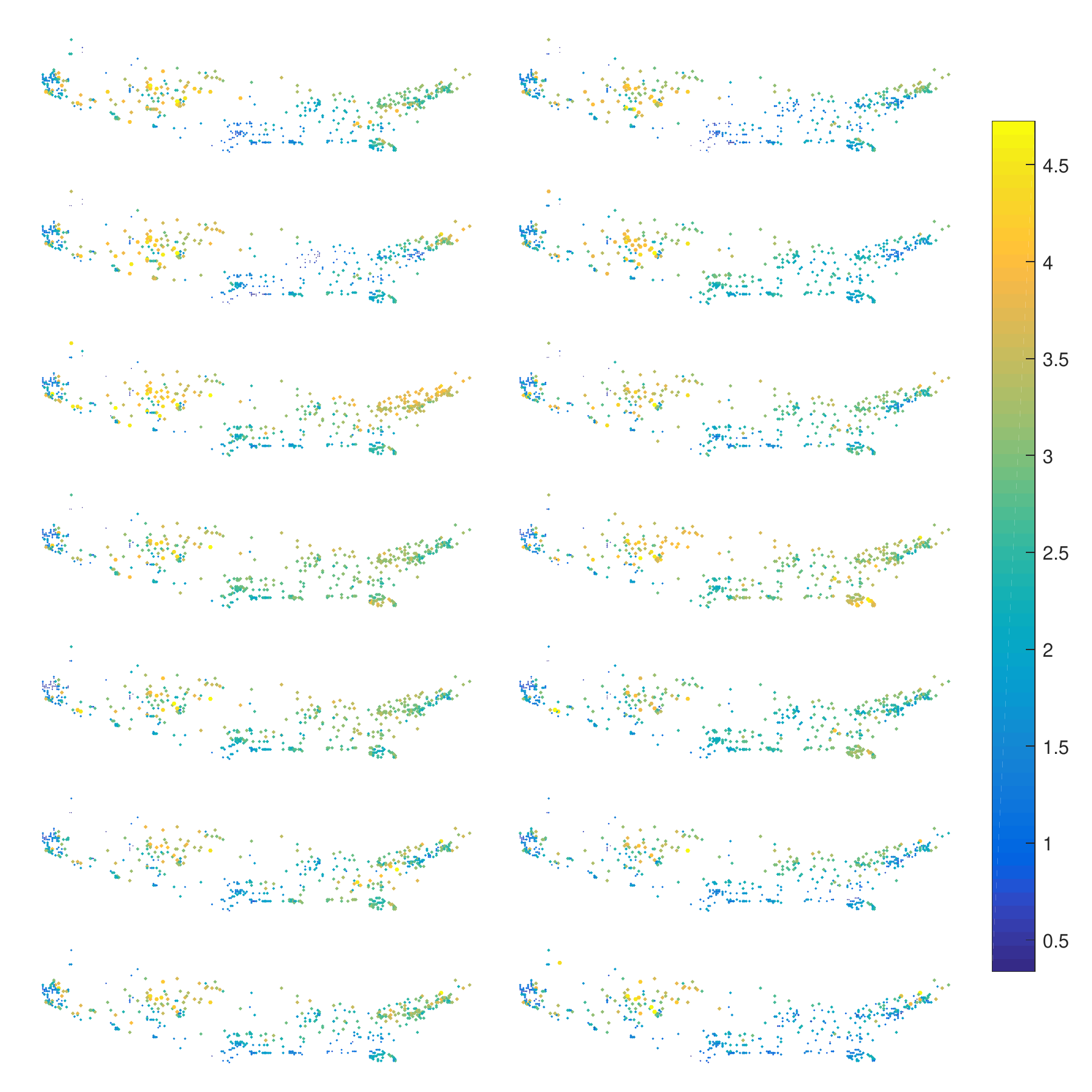} %
		\caption{Ozone daily average levels measured over the conterminous USA for 14 consecutive days in 2015 starting on January 1, 2015.
The colorbar scale corresponds to ozone concentration levels multiplied by 100. The actual ozone levels (in ppm) are obtained by multiplying the colorbar scale
with  $10^{-2}$.  According to EPA the air quality standard for ozone is 0.075 ppm, averaged over eight hours.}
		\label{fig:data_scatter}
	\end{center}
\end{figure}

\begin{figure}[htbp]
	\begin{center}
		\includegraphics[width=0.8\textwidth]{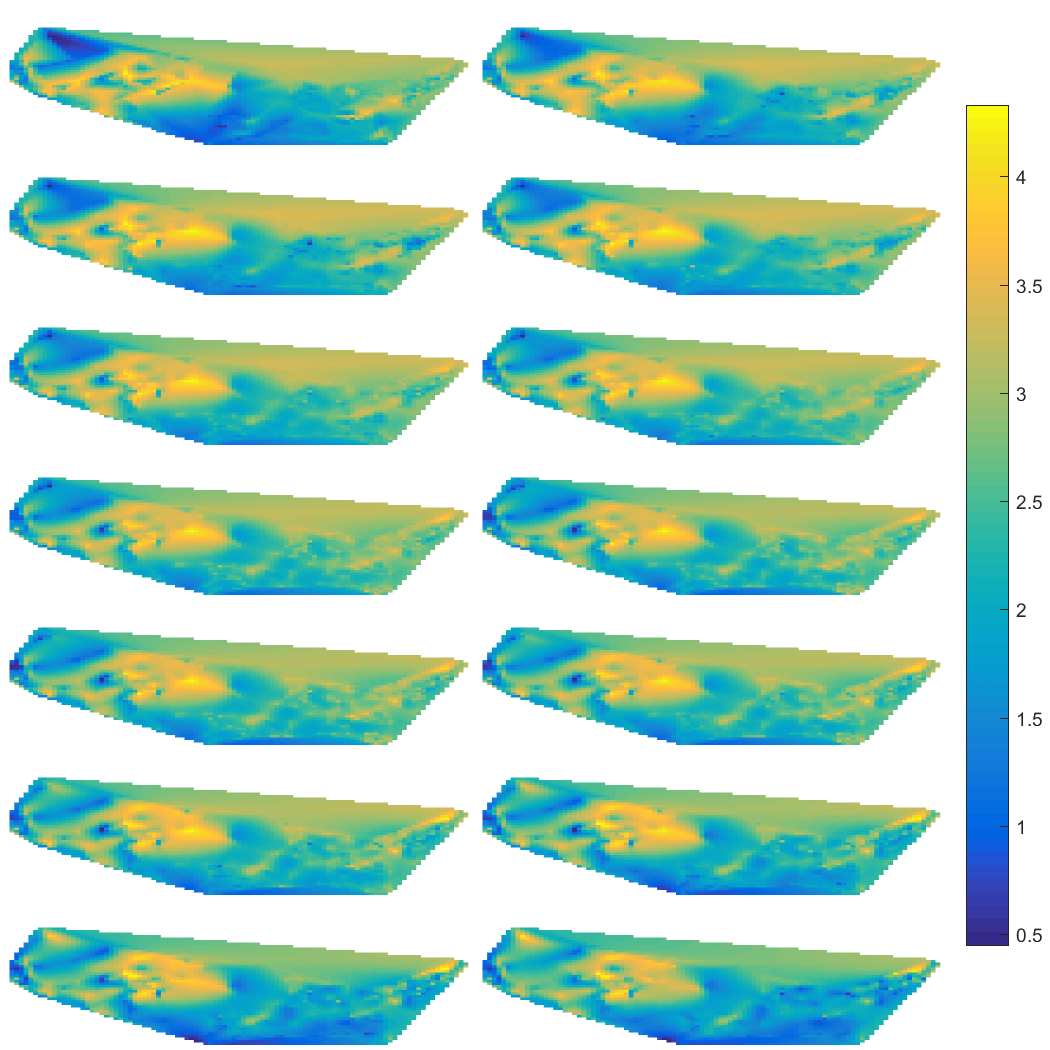} %
		\caption{Maps of ozone concentration  on a $100 \times 50$ map grid based on natural neighbor interpolation of the
data shown in Fig.~\ref{fig:data_scatter}.}
		\label{fig:maps_interp}
	\end{center}
\end{figure}

The  spatial distribution of the locations and the ozone levels are shown in Fig.~\ref{fig:data_scatter}.
Maps of approximate ozone concentrations are obtained by means of  natural neighbor interpolation~\cite{Ledoux05}
and they are displayed in Fig.~\ref{fig:maps_interp}.
Based on the visual inspection of these maps we postulate a quadratic spatial trend function. The latter is modeled by the second degree polynomial
\beq
\label{eq:trend}
m(\bfs) = c_{0} + c_{1}s_{1} + c_{2} s_{2} + c_{1,1} s_{1}^{2} + c_{2,2} s_{2}^2 + c_{1,2} s_{1}\,s_{2},
\eeq
where $\bfs = (s_{1}, s_{2})$ are the location vectors on the plane. The spatial coordinates of the station locations  have been
converted to the World Geodetic System 1984 (WGS84) from the initial longitude-latitude format.
Subsequently, the spatial coordinates have been  normalized by dividing with $10^4$m.

The coefficients of the optimal (in the least squares sense) trend model are given in Table~\ref{tab:trend}.
Most of the coefficients have negative values even though the ozone concentrations are positive. The sign of the coefficients is due to the preponderance of
negative values among the spatial coordinates of the stations.

\begin{table}
  \centering
  {\footnotesize \begin{tabular}{cccccccc}
  \hline
  $c_{0}$ & $c_{1}$ &  $c_{2}$ & $c_{1,1}$ & $c_{2,2}$ & $c_{1,2}$ & $R$ & $p$  \\
  \hline
  $-4.44$ & $-9.25\, 10^{-4}$ & $-0.035$ & $-1.56\, 10^{-5}$ & $-4.3\, 10^{-5}$ & $-2.49\, 10^{-6}$ & 0.455 & $1.6 \, 10^{-29}$  \\
  \hline
\end{tabular}}
  \caption{Coefficients of the quadratic trend regression model~\eqref{eq:trend}, as well as the value of the correlation coefficient
  $R$ between the trend and the data, and the $p$ value of the trend model. }
  \label{tab:trend}
\end{table}

\begin{figure}[htbp]
	\begin{center}
		\includegraphics[width=0.8\textwidth]{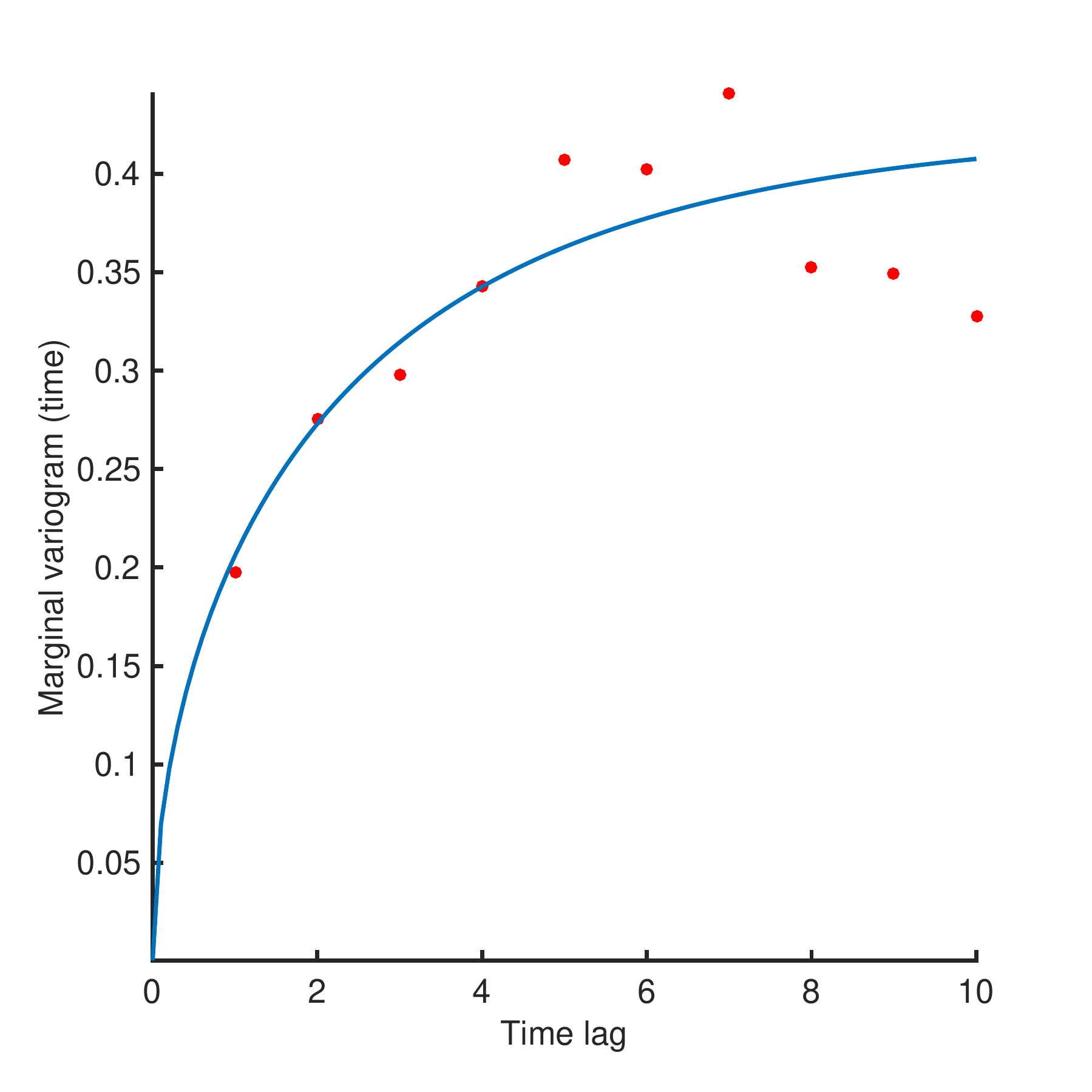} %
		\caption{Marginal temporal variogram estimated from the data (circles) and best fit to the theoretical STSLR temporal model~\eqref{eq:vari-time}.
The time lag is measured in terms of days. The vertical axis is measured in $\mathrm{ppm}^2 \times 10^{4}$.}
		\label{fig:vario_time}
	\end{center}
\end{figure}

\begin{figure}[htbp]
	\begin{center}
		\includegraphics[width=0.8\textwidth]{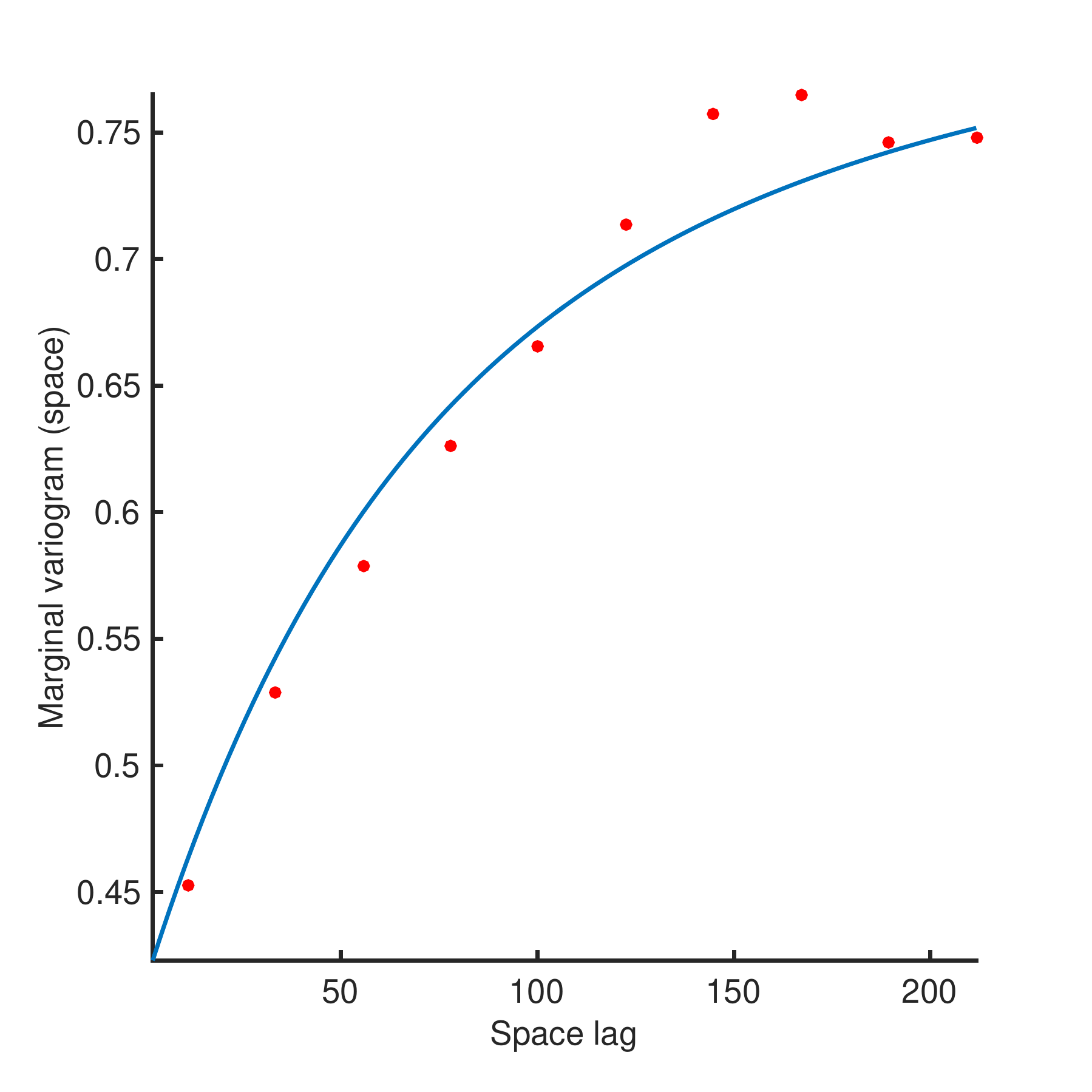} %
		\caption{Marginal spatial variogram estimated from the data (circles) and best fit to the theoretical STSLR spatial model~\eqref{eq:vari-space}.
The actual spatial lag is obtained by multiplying the horizontal axis with $10^4$m. The units of the vertical axis aren $\mathrm{ppm}^2 \times 10^{4}$.}
		\label{fig:vario_space}
	\end{center}
\end{figure}

We use the detrended ozone data to estimate the marginal variograms according to~\eqref{eq:temporal-vario-estim} and~\eqref{eq:spatial-vario-estim}.
The empirical temporal marginal variogram and its fit to the theoretical STSLR marginal function~\eqref{eq:vari-time} is shown in Fig.~\ref{fig:vario_time}.
The empirical omnidirectional spatial marginal variogram and the respective fit to the theoretical STSLR marginal function~\eqref{eq:vari-space}
is shown in Fig.~\ref{fig:vario_space}. The estimated parameters for the STSLR variogram model are as follows:
$\eta_{0} = 0.7924$ $\mathrm{ppm}^2 \times 10^{4}$, $\la =1.07$ (dimensionless flexibility),
 $\xi=45.49$ in normalized units (equivalently $\xi \approx 450$km), $\tau_c=4.70$ days.

The empirical spatial marginal variogram includes a nugget term
with an estimated variance  $c_{0}=0.4125$, while the nugget is negligible in the case of the temporal marginal variogram.
The characteristic length $\xi$ may seem high. However, it agrees with the smooth maps produced by means of natural neighbor interpolation in Fig.~\ref{fig:maps_interp}.
The average distance between each station and its closest neighbor is $3.28$ (32.8km) while the maximum distance between nearest neighbors is
26.74 (267.4km) in normalized units (km).
This relatively large separation between stations accounts for the loss of spatial resolution during ozone sampling.
This effect is also apparent in the finite
nugget of the marginal spatial variogram.

\begin{figure}[htbp]
	\begin{center}
		\includegraphics[width=0.8\textwidth]{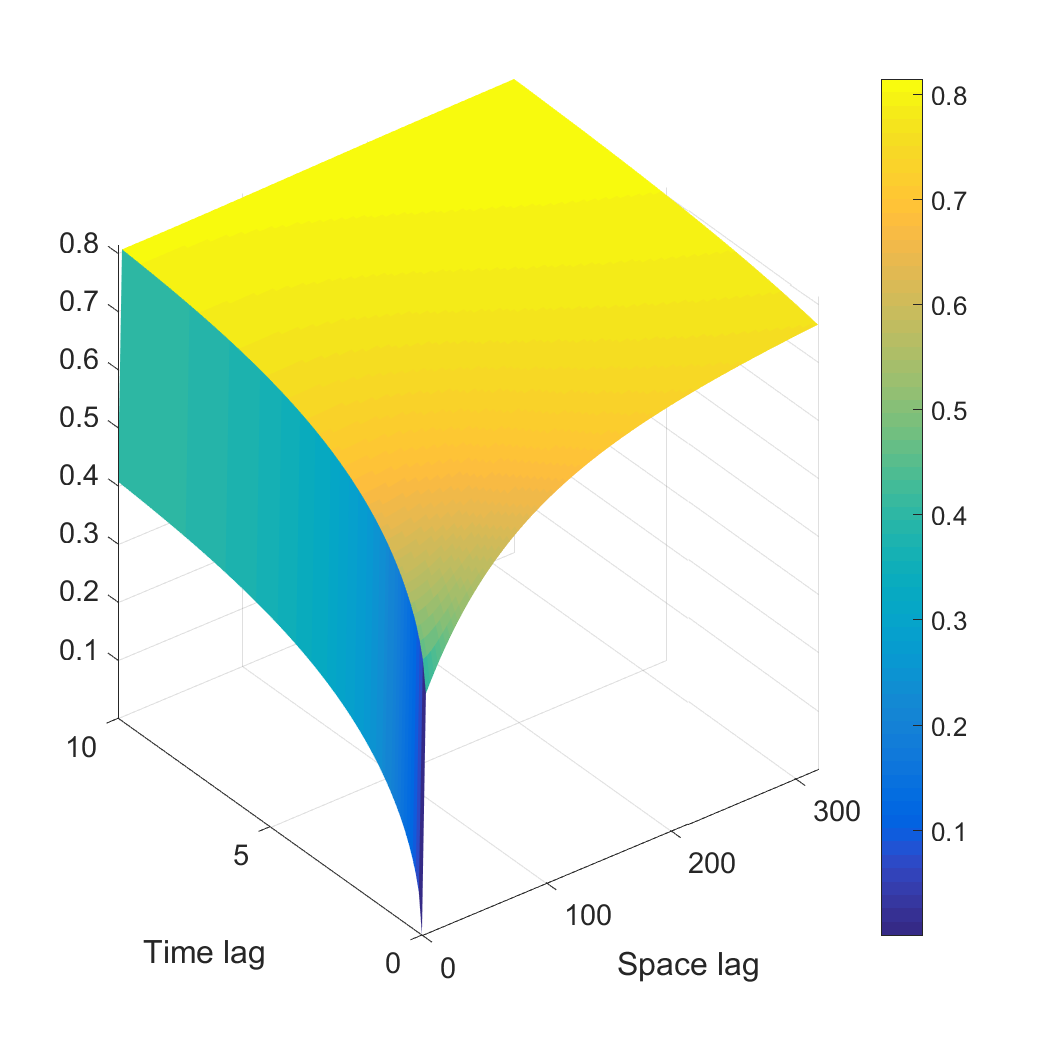} %
		\caption{Space-time variogram model for the ozone data which consists of the STSLR space-time model~\eqref{eq:vario-sli} and a spatial nugget term.
The time lag is measured in days, and the spatial lag has been normalized by dividing with $10^4$m. The vertical axis is measured in $\mathrm{ppm}^2 \times 10^{4}$.}
		\label{fig:vario_model}
	\end{center}
\end{figure}

The non-separable space-time STSLR variogram~\eqref{eq:vario-sli} based on the
estimated optimal parameters for the ozone data is shown in Fig.~\ref{fig:vario_model}  including  the
spatial nugget component. The projection on the plane $\rr=0$ represents the marginal STSLR time variogram  without a nugget.
The marginal STSLR time variogram is estimated by averaging the temporal variograms of the ozone time series at each location, and thus it
does not incorporate spatial variability; the latter appears as  a discontinuity for $\rr>0$. The nugget term explains the
difference between the sills of the marginal time and space variograms in the plot, since the sills of the STSLR marginal variograms are
identical by construction.

%
%
%
%
%

\section{Discussion and Conclusions}
\label{sec:conclu}
There is an ongoing interest in the development of flexible and realistic spatiotemporal covariance functions.
We explore the generation of non-separable space-time covariance functions based on the solutions of respective equations of motion.
The latter are partial differential equations that follow from the theory of linear response which describes the dynamic fluctuations
of random fields around an equilibrium point.

The covariance equation of motion~\eqref{eq:cor-eom}  is based on an equilibrium
``energy function'' determined by the squares of the fluctuating field, its gradient, and its curvature.
Combining the explicit solution in one spatial dimension of the covariance equation of motion~\eqref{eq:cor-eom} with the turning bands method, we obtain
the novel STSLR space-time covariance~\eqref{eq:new-st} which comprises a combination of exponential factors,
error functions, and algebraic powers. The STSLR covariance  is a non-separable, space-time stationary, and spatially isotropic
function which does not belong in the Gneiting class. In addition, the STSLR space-time covariance is everywhere continuous but non-differentiable at zero space and/or time lag.

The STSLR space-time covariance includes four parameters:
an overall scale factor, $\eo$, a characteristic time, $\tau_c$, a characteristic length, $\xi$, and a flexibility coefficient, $\la$, which is
related to the ``resistance'' of the spatial fluctuations to spatial gradients. Larger values of $\la$ imply that the random field is
more likely to admit higher gradients than for lower $\la$.
In addition, it is possible to incorporate geometric anisotropy  by means of rotation and scaling transformations of the
spatial lag vector in~\eqref{eq:new-st}, e.g.~\cite{dth08}. Future research will focus on developing efficient interpolation and simulation schemes based on
the STSLR covariance function as well as comparisons with other covariance models.

A more general question is whether the linear response theory can be extended to include solutions of hyperbolic type.
Another interesting problem is whether a tractable solution of~\eqref{eq:cor-eom} can be derived  in three spatial dimensions, in order to
avoid the application of the turning bands transform. A solution has been obtained at the zero-curvature limit, but this
function diverges at $\rr=0$~\cite{dth15b}. A finite curvature term is needed in order to overcome the divergence.
However, the integral of the spectral density for finite curvature is not analytically tractable, which means that a closed-form expression can at best be derived
as a series expansion. Such an expansion may lead to  more flexible parametric covariance forms.


\section*{Acknowledgments}
{The research presented in this manuscript was partly funded by the project SPARTA 1591: ``Development of
Space-Time Random Fields based on Local Interaction Models and Applications in the Processing of
Spatiotemporal Datasets''. The project SPARTA was implemented under the ``ARISTEIA'' Action  and was co-funded by the European Social Fund
 and National Resources.}
%
\bibliographystyle{unsrt}

\bibliography{st}

\end{document}